\documentclass{article}
\usepackage{tikz}
\usepackage[affil-it]{authblk}
\usepackage{amsmath, amssymb}
\usepackage{enumitem}
\usepackage[normalem]{ulem}
\usepackage{enumitem}
\usepackage{soul}
\usepackage{float}

\begin{document}
\newcommand{\Seselja}{\v{S}e\v{s}elja}
\newcommand{\dan}[1]{{\color{blue}#1}}
\newcommand{\rev}[1]{{\color{red}#1}}
\newcommand{\am}[1]{{\color{olive}#1}}
\newcommand{\chr}[1]{{\color{blue}#1}}
\newcommand{\dd}[1]{{\color{red}#1}}
\newcommand{\new}[1]{{\color{red}#1}}
\renewcommand{\dd}[1]{#1}
\renewcommand{\am}[1]{#1}
\renewcommand{\rev}[1]{#1}
\renewcommand{\new}[1]{#1}
\renewcommand{\chr}[1]{#1}

\title{An argumentative agent-based model of scientific inquiry}

\author{AnneMarie Borg}
\affil{Institute for Philosophy II, Ruhr-University Bochum}
\author{Daniel Frey}
\affil{Heidelberg University}
\author{Dunja \v{S}e\v{s}elja}
\affil{Institute for Philosophy II, Ruhr-University Bochum\\Center for Logic and Philosophy of Science, Ghent University}
\author{Christian Stra\ss er}
\affil{Institute for Philosophy II, Ruhr-University Bochum\\Center for Logic and Philosophy of Science, Ghent University}

\maketitle

\begin{abstract}
  In this paper we present an agent-based model (ABM) of scientific
  inquiry aimed at investigating how different social networks impact
  the efficiency of scientists in acquiring knowledge. As such, the
  ABM is a computational tool for tackling issues in the domain of
  scientific methodology and science policy. In contrast to existing
  ABMs of science, our model aims to represent the argumentative
  dynamics that underlies scientific practice. To this end we employ
  abstract argumentation theory as the core design feature of the
  model. This helps to avoid a number of problematic idealizations
  which are present in other ABMs of science and which impede their
  relevance for actual scientific practice.
\end{abstract}

\section{Introduction}
\label{sec:intro}

In this paper we present a computational tool for tackling issues in the domain of scientific methodology and science policy, which concern social aspects of scientific inquiry and the division of cognitive labor. Recent approaches to these questions
have utilized
     agent-based models (ABMs) (e.g.
     \cite{zollman2007communication,zollman2010epistemic},
     \cite{weisberg2009epistemic}, \cite{douven2010simulating},
     \cite{thoma2015epistemic}, etc). One of the advantages of such
     formal approaches is that they avoid hasty generalizations
     resulting from the traditional case-study approach since they
     employ an experimental method in which a number of relevant
     variables can be controlled during the run of simulations.
     However, the main pitfall of ABMs is that they frequently suffer
     from a high degree of idealization and simplification, which may
     sometimes impede the relevance of the results for the context of
     actual scientific inquiry
     \cite{payette2012agent,thoma2015epistemic}.

     One such idealization concerns the question how we represent
     evaluations on the basis of which scientists choose paths
     to pursue. Ever since Laudan introduced the notion of the context
     of pursuit \cite{Laudan77}, many have argued that the assessment
     of one's research direction has to be distinguished from the
     epistemic justification of one's beliefs (e.g.\
     \cite{Nickles06,Seselja_Kosolosky_12}). While in some ABMs these
     two types of evaluations have been conflated (e.g.\
     \cite{zollman2007communication,zollman2010epistemic}), in others
     the former type of assessment has been represented in an
     oversimplified way (e.g.\ \cite{weisberg2009epistemic}). In both
     cases the way in which scientists respond to the information
     which they receive from their peers, and in view of which they
     update their pursuit-related attitudes, has not been modeled
     according to the idea of \emph{heuristic appraisal}
     \cite{Nickles06}. This idea assumes that when scientists face
     possible anomalies in their research programs, such anomalies
     trigger a search for counterarguments that could turn apparent
     refutations into confirmatory instances \cite{Lakatos78}.

     This points to another problematic idealization frequently
     present in ABMs, namely the fact that the interaction among
     scientists is represented as a simple update in view of received
     information. While some models have tried to relax this
     idealization by representing agents as ``trusting'' others only
     if they have sufficiently similar views (e.g.\
     \cite{hegselmann2005opinion}), by introducing ``noise'' in the
     received information \cite{douven2010simulating}, by assigning
     different weights to the opinions of agents
     \cite{riegler2009extending}, or by assigning different epistemic
     systems to agents \cite{de2012peer}, these adjustments do not
     capture the argumentative nature of scientific interaction, such
     as the above mentioned search for a counterargument in face of an
     attack.

In this paper we offer a novel approach to the agent-based modeling of
scientific inquiry, which is based on argumentation, and which aims to
soften the above mentioned idealizations. In contrast to most other
ABMs of science, our model is based on the idea that an essential
component of scientific inquiry is an argumentative dynamics between
scientists. To this end, we employ abstract argumentation
framework\am{s} as one of the design features of our ABM (previously
shown fruitful for the modeling of scientific debates in
\cite{Seselja_Strasser_11} \rev{and employed in an ABM of social
  behavior in \cite{gabbriellini2014new}}).

The presented version of the model is designed to investigate how
different social networks impact the efficiency of scientists in
discovering the best of the pursued scientific theories. This question
has previously been tackled in
\cite{zollman2007communication,zollman2010epistemic} whose ABMs
suggest that information sharing may sometimes be counter-productive
for the efficiency in knowledge acquisition. This is due to the fact
that scientists may initially get misleading results, suggesting that
theory $T_1$ is better than theory $T_2$, while from an objective
point of view it is the other way around. If such misleading
information is shared widely, the whole scientific community may start
pursuing $T_1$ and abandon $T_2$. However, recent discussion in
\cite{rosenstock2016epistemic} has shown that the results by Zollman
\cite{zollman2007communication,zollman2010epistemic} regarding the
harmful effects of increased interaction are not robust under some
minor changes of the relevant parameters. Nevertheless, the authors
also point out that similar results have been obtained by a
structurally different model, namely the one by Grim
\cite{grim2009threshold}, which leaves the question of the epistemic
benefits of scientific interaction open. Moreover, both Zollman's and
Grim's models have been criticized as suffering from various
problematic idealizations in \cite{seselja_ABMs_interaction}, which
impede the relevance of their results for actual scientific practice.
One of these simplifications is the above mentioned representation of
scientific interaction as a simple update in view of new evidence.
The aim of our model is to tackle the same question, while
representing scientific interaction in a more adequate way.

The paper is structured as follows. In Section \ref{the-model} we
introduce the idea underlying our ABM and its main features. In
Section \ref{sec:findings-discussion} we present the central findings
of the presented version of our model. In Section \ref{sec:discussion}
we compare our findings with those obtained by other ABMs of science
and discuss idealizations present in our current model, which should
be taken into consideration when assessing the relevance of our
results for the actual scientific practice. In Section
\ref{sec:outlook-conclusion} we suggest ideas for further enhancements
of this ABM.


\section{The model}
\label{the-model}
The aim of our ABM is to represent scientists
engaged in a scientific inquiry with the goal of finding the best of
the given rivaling theories, where they occasionally exchange
arguments with other scientists. In this section we will explain the
central elements of our model, namely, the landscape which agents
explore, the behavior of agents and the notion of social networks
employed in our simulations.\footnote{The source code is available at\\
  https://github.com/g4v4g4i/ArgABM/tree/AppArg2017-submission.}

\def\disc{\hookrightarrow}
\def\att{\leadsto}

\subsection{The landscape}
\label{sec:landscape}


Agents, representing scientists, move along an \emph{argumentative
  landscape}. The argumentative landscape, which represents rivaling
theories in a given scientific domain, is based on a dynamic abstract
argumentation framework. Let us explain what `abstract' and `dynamic'
mean here.


An abstract argumentation framework (AF), introduced by Dung in
\cite{journals/jlp/Dung95} 
is a formal framework consisting of a set of abstract entities
$\mathcal{A}$ representing \emph{arguments} and an \emph{attack
  relation} ${\att}$ over $\mathcal{A}$. Similarly, the framework
underlying our model consists of a set of arguments and an attack
relation over this set. In addition to attacking each other, arguments
may also be connected by a \emph{discovery relation} $\disc$. The
latter represents the path which scientists have to take in order to
discover different parts of the given theory, i.e., some argument $b$
in the given theory can only be discovered after an argument $a$ has
been discovered for which $a \disc b$.

A scientific theory is represented as a conflict-free set of arguments (i.e.\ no argument in the theory attacks an argument in the same theory), connected by discovery relations, resulting in a tree-like graph. Formally, an argumentative landscape is given by a triple $\langle \mathcal{A}, \att, \disc\rangle$ where $\mathcal{A} = \langle \mathcal{A}_1, \ldots, \mathcal{A}_m \rangle$ is partitioned in $m$ many theories $T_i = \langle \mathcal{A}_i, a_i, \disc\rangle$ which are trees with $a_i \in \mathcal{A}_i$ as a root and 
$${\att} \subseteq \bigcup_{\substack{1 \le i, j \le m \\ i \neq j}} (\mathcal{A}_i \times \mathcal{A}_j) \quad \mbox{ and } \quad {\disc} \subseteq \bigcup_{1 \le i \le m} (\mathcal{A}_i \times \mathcal{A}_i).$$
Specifying $\att$ like this  ensures that the theories are conflict-free.

The abstractness of the framework concerns all of its elements. Instead of providing the concrete content and structure of the given arguments, we represent them as abstract entities. Similarly, we do not reveal the concrete nature of the attack or the discovery relation.

The framework is dynamic in the sense that agents gradually discover
arguments, as well as attack and discovery relations between them.
Given the abstract nature of arguments, we interpret them as
hypotheses which scientists investigate, occasionally encountering
defeating evidence, represented by attacks from other arguments, and
then attempting to find defending arguments for the attacked
hypothesis $a$ \dd{(i.e., to find arguments in the same theory
  attacking arguments from other theories that attack $a$).} This
dynamic aspect is implemented by associating arguments with their
\emph{degree of exploration} for an agent at a given time point of a
run of the simulation: for each agent $\mathsf{ag}$ and each argument
$a \in \mathcal{A}$,
$\mathfrak{expl}(a, \mathsf{ag}) \in \{0, \ldots, 6\}$ where $0$
indicates that the argument is unknown to $\mathsf{ag}$ and $6$
indicates that the argument is fully explored and cannot be further
explored.\footnote{\new{Our model is round-based (more on that in Section \ref{sec:basic-behav-agents}). Each round may be interpreted as one research day. Since each of
    the 6 levels of an argument is explored in 5 rounds, each argument
    represents a hypothesis that needs 30 research days to be fully
    investigated.}} In view of this agents have subjective and limited
insights into the structure of the landscape. Whether an attack or
discovery relation between two arguments $a$ and $a'$ is visible to an
agent $\mathsf{ag}$ depends on the degree of exploration
$\mathfrak{expl}(a, \mathsf{ag})$: the higher
$\mathfrak{expl}(a, \mathsf{ag})$ is, the more relation[s] between $a$
and other arguments will be visible
(additionally agents may learn about the landscape by communicating with other agents, see Section \ref{sec:social-netw}).

\subsection{Basic behavior of agents}
\label{sec:basic-behav-agents}

The model is round-based and each round agents perform actions which are among the following:

\begin{enumerate}[itemsep=0pt,topsep=1pt]
\item exploring a single argument, thereby gradually discovering possible attacks (on it, and from it to an argument that belongs to another theory) as well as discovery relations to neighboring arguments;
\item moving to a neighboring argument along the discovery relation within the same theory;
\item moving to an argument of a rivaling theory.
\end{enumerate}

While agents start the run of the simulation at the root of a given theory, they will gradually discover more and more of the argumentative landscape. 
This way each turn an agent operates on her own (subjective) fragment of the landscape, which consists of her discovered arguments which are explored by her to a specific degree, and her discovered (attack and discovery) relations between the arguments.

In order to decide whether to work on the current theory (items 1 and
2 above), or whether to better start working on an alternative theory
(item 3) agents are equipped with the ability to evaluate theories.
Every few rounds agents apply an evaluative procedure, which \chr{is
  based on the \emph{degree of defensibility} of each of the given
  theories. A theory has degree of defensibility $n$ if it has $n$
  defended arguments where an argument $a$ is defended in the theory
  if each attacker $b$ from another theory is itself attacked by some
  argument $c$ in the current theory. Agents decide to move to a
  rivaling theory if the degree of defensibility of their current
  theory is below a relative threshold compared to the theory with the
  highest degree of defensibility, i.e., the theory with the most
  defended
  arguments.} 

To make a decision between options 1 and 2, each agent employs the
following heuristic: at every time step she considers all arguments in
her direct neighborhood (relative to the discovery relation) that
could possibly be the next ones to work on. With a certain probability
she will then move to one of these arguments, or alternatively keep on
exploring her current argument. In case the argument she's currently
at is fully explored, she will try to move on to a next neighboring
argument, and if such an argument isn't visible (e.g., if she has
reached the end of a branch of her theory) she will try to move to
\new{the parent argument, or in case it is fully explored}, she will
move to another not fully explored argument in the same theory.

The decision making of agents also includes some prospective
considerations. If during her exploration an agent discovers an attack
on the argument $a$ she is currently investigating, she will attempt
to discover a defense for it. A defending argument may be found among
the visible neighboring arguments of $a$ (relative to the discovery
relation). In case $a$ is attacked by arguments $b_1, \ldots, b_n$, an
agent $\mathsf{ag}$ working on $a$ will `see' outgoing attack arrows
$a' \att b_i$ (where $1 \le i \le n$) from an already discovered
\new{child argument} $a'$ of $a$, even in cases where $a' \att b_i$ is
not yet discovered by $\mathsf{ag}$ since $a'$ is insufficiently
explored by $\mathsf{ag}$.\footnote{In such cases, where an attack
  relation is merely `seen' but not yet discovered by an agent, it is
  not yet considered as a defense when the respective theory is
  evaluated by the agent.} This way our agent knows that exploring
$a'$ may help in defending $a$. If no potential defender of $a$ is
visible she keeps on exploring $a$ in the hope of discovering new
neighbors and thus new potential defenders.

This idea corresponds to the situation in which a scientist discovers
a problem in her hypothesis and attempts to find a way to resolve it.
While she may not have a solution ready at hand, she may have a method
for finding such a solution (for example, going back to the laboratory
and conducting some new experiments).\footnote{Such a heuristic
  response to apparent refutations belongs to what Lakatos has dubbed
  the \emph{negative heuristics} of a research program
  \cite{Lakatos78}.}


\subsection{Social networks}
\label{sec:social-netw}

An agent discovers the argumentative landscape by investigating
arguments (as described in Section \ref{sec:basic-behav-agents}) or by
means of exchanging information about the landscape with other agents.
We will now discuss the later aspect. %
As mentioned above, the presented version of the model is designed to
investigate how different social networks impact the efficiency of
scientists in discovering the best
theory. 
In contrast to other ABMs employing the idea of social networks (e.g.
\cite{zollman2007communication,zollman2010epistemic,grim2009threshold}),
which represent connectivity only in view of different types of graphs
that connect agents, \new{we distinguish between two types of social
  networks}.

First, our agents are divided into \emph{collaborative networks} that
may consist of individuals working on the same theory (homogeneous
groups), or of individuals working on different theories
(heterogeneous groups). While each agent gathers information (i.e.\
the attack and discovery relations between arguments) on her own,
every five \new{steps} this information is shared with all other
agents forming the same collaborative network.

Second, besides sharing information with agents from the same network, \new{every five steps} each agent shares information with agents from other collaborative
networks with a given \emph{probability of information sharing} that is determined before the run of the simulation.\footnote{While agents share their full subjective knowledge
  within their respective collaborative networks, the information
  which they share with agents from other networks concerns
  \dd{recently obtained knowledge of the theory} which they are
  currently exploring. This corresponds to a situation in which a
  scientist writes a paper that presents arguments for and/or against
  hypotheses regarding the theory she is currently pursuing.}
\new{This way the agents form ad-hoc and random networks with agents from other research collaborations. A higher probability of information sharing leads to a
  higher degree of interaction among
  agents}.

Finally, we represent \emph{reliable} and \emph{biased} scientists by
allowing for different approaches to the sharing of information
between networks. A reliable agent shares all the information she has
gathered during her exploration of the current theory, while a biased
agent does not share the information regarding the discovered attacks
on her current theory.

Agents share information either in a unidirectional (an agent sends
information to another agent) or a bidirectional way (agents exchange
information one with another). Moreover, our model takes into account
the fact that receiving information is time costly:
when an agent receives information, she will not explore the argument
on which she is standing nor move. This corresponds to the idea that
scientists need to invest time when reading papers by other
scientists, which they would otherwise devote to their own research.

\section{The main findings}
\label{sec:findings-discussion}

In this section we will first specify the parameters used in the
simulations and then present our main results.

\subsection{Parameters used in simulations}
\label{sec:parameters}

We have run the simulation 100 times with 10, 20, 30, 40, 70 and 100 agents
by varying the following settings:
\begin{enumerate}[itemsep=0pt,topsep=0pt]
\item different probabilities of an agent communicating with agents from other collaborative networks, namely: 0, 0.3, 0.5, and 1;
\item different types of collaborative networks, namely: homogeneous and heterogeneous ones;
\item different approaches to communicating, namely: reliable and
  biased agents;
\item two different landscapes, one representing two theories, and one
  representing three theories.\footnote{Some other parameters are as
    follows. Theories are modeled as trees of depth 3, where each
    argument (except for the final leaves) has 4 child-arguments. The
    extent to which each of the theories is attacked is specified in
    the interface of the model in terms of the probability that each
    of the arguments of the given theory is attacked. Here we have
    opted for 0.3 probability since we wish to represent a situation
    in which rivaling theories are not completely problematic (as
    would be e.g.\ pseudo-scientific theories).}
\end{enumerate}


The program runs until each agent is on a fully explored theory, since at this point the agents are locked in the
given theories and no further information is available, in view of
which they would move to another theory.
Our main research question is how efficient agents are in each of
the above listed scenarios, where efficiency is assessed in terms of
the success of agents in acquiring knowledge, as well as in terms of
the time needed for the run to be completed.

In this version of the model we have defined success in the following
way: if, at the end of the run, \chr{there is no theory for which the number of agents working on it is greater than the number of agents working on the objectively best theory,}
the run is considered
successful. In contrast to some other ABMs of science, which define
success in terms of convergence of all agents onto the best theory
(e.g.\ in \cite{zollman2007communication,zollman2010epistemic}), our
notion is obviously weaker.
Our choice was motivated by a pluralist view on scientific inquiry,
according to which, a parallel existence of rivaling scientific
theories is epistemically and heuristically beneficial for the goals
of a scientific community (e.g.\
\cite{Longino_02,kitcher2011science,Chang_12}). A primary epistemic
concern is thus not the convergence of all scientists onto the same
theory, but rather assuring that the best theory is among the most
actively investigated ones.

\subsection{Preliminary results}
\label{sec:results}

\begin{figure}[t]
\centering
\includegraphics[width=0.8\linewidth]{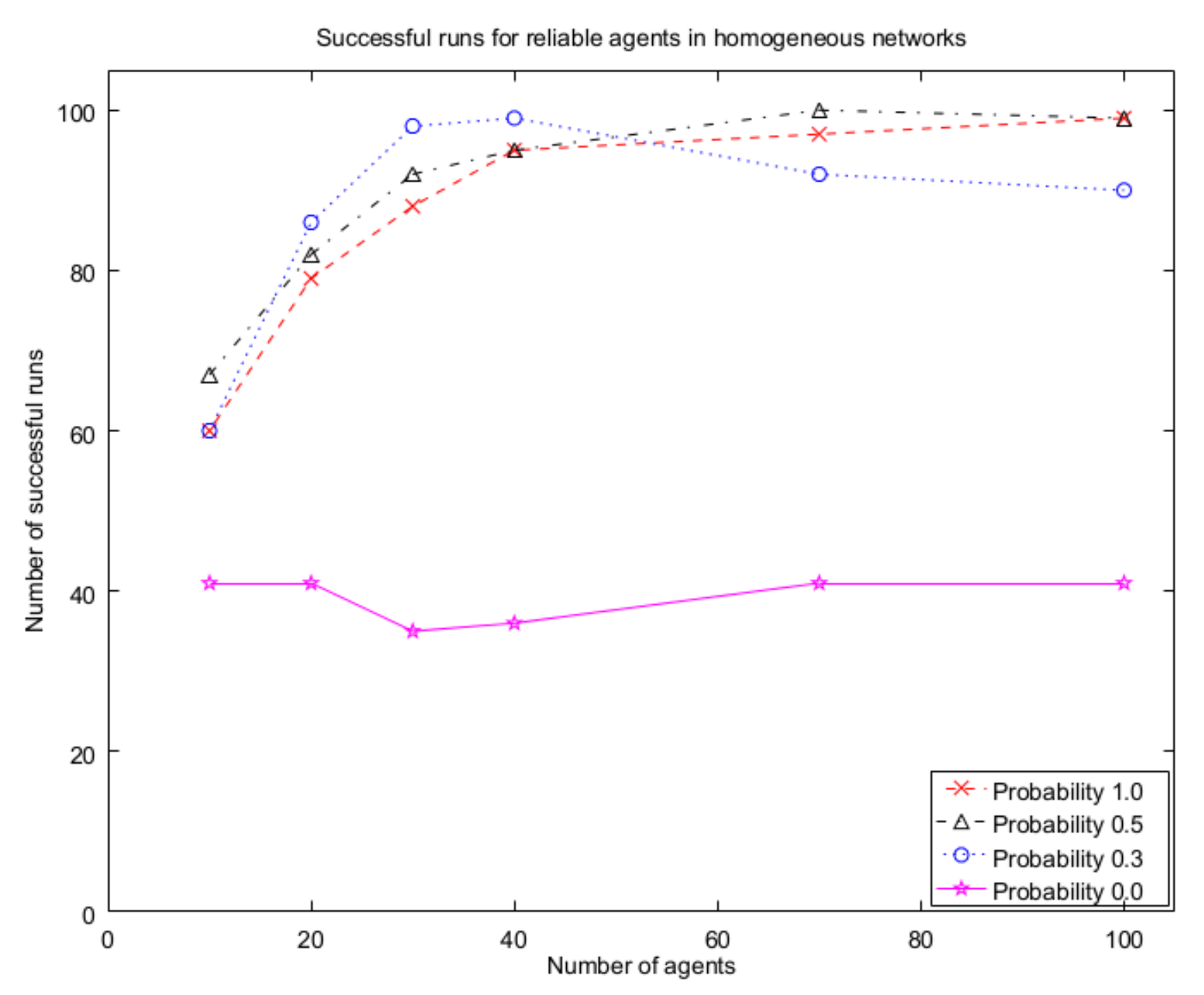}
\caption{Success, reliable, homogeneous groups}
\label{fig:sucrelhomog}
\end{figure}

\begin{figure}[p]
\centering
\includegraphics[width=0.8\linewidth]{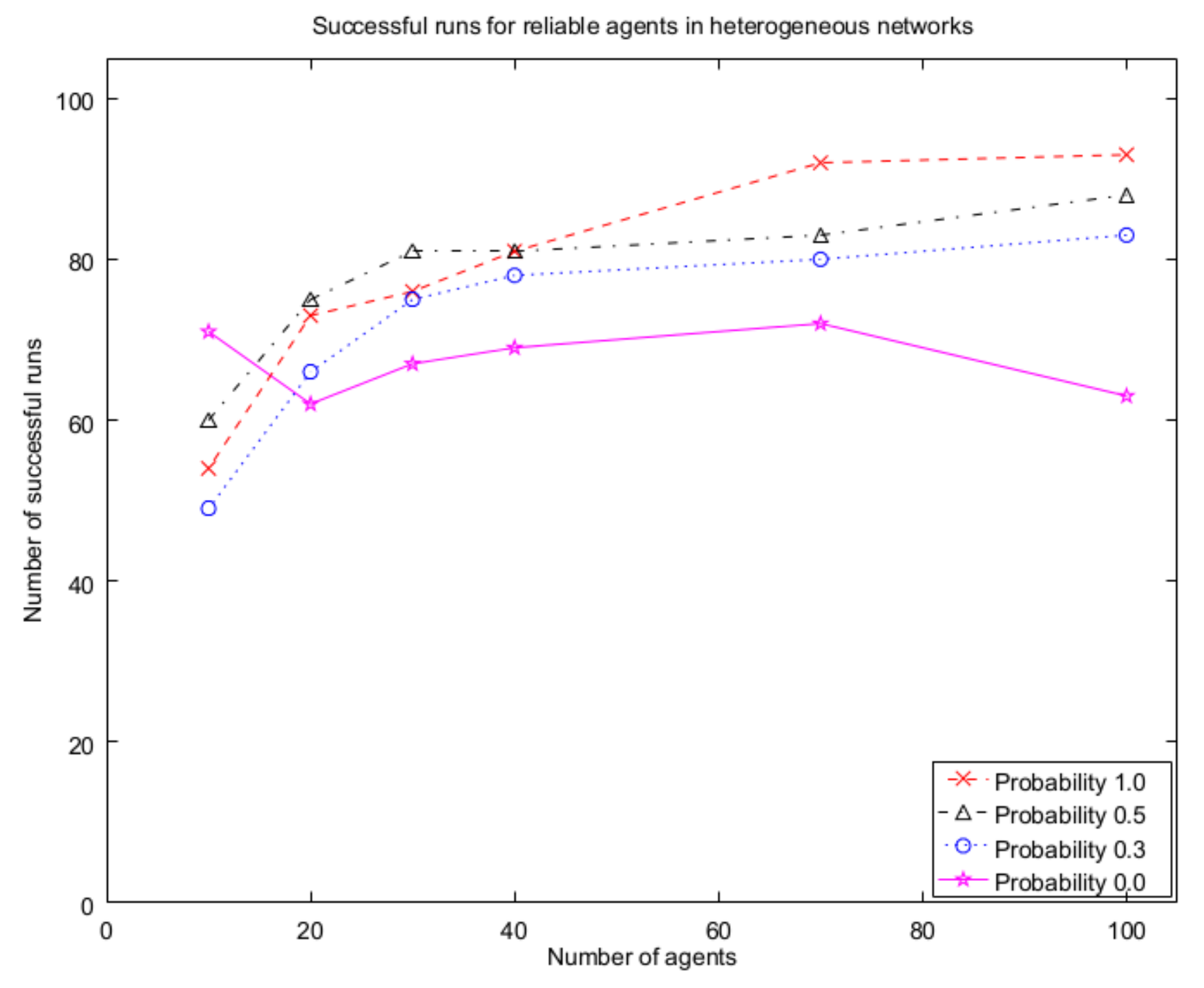}
\caption{Success, reliable, heterogeneous groups}
\label{fig:sucrelheterog}
\end{figure}

\begin{figure}[p]
\centering
\includegraphics[width=0.8\linewidth]{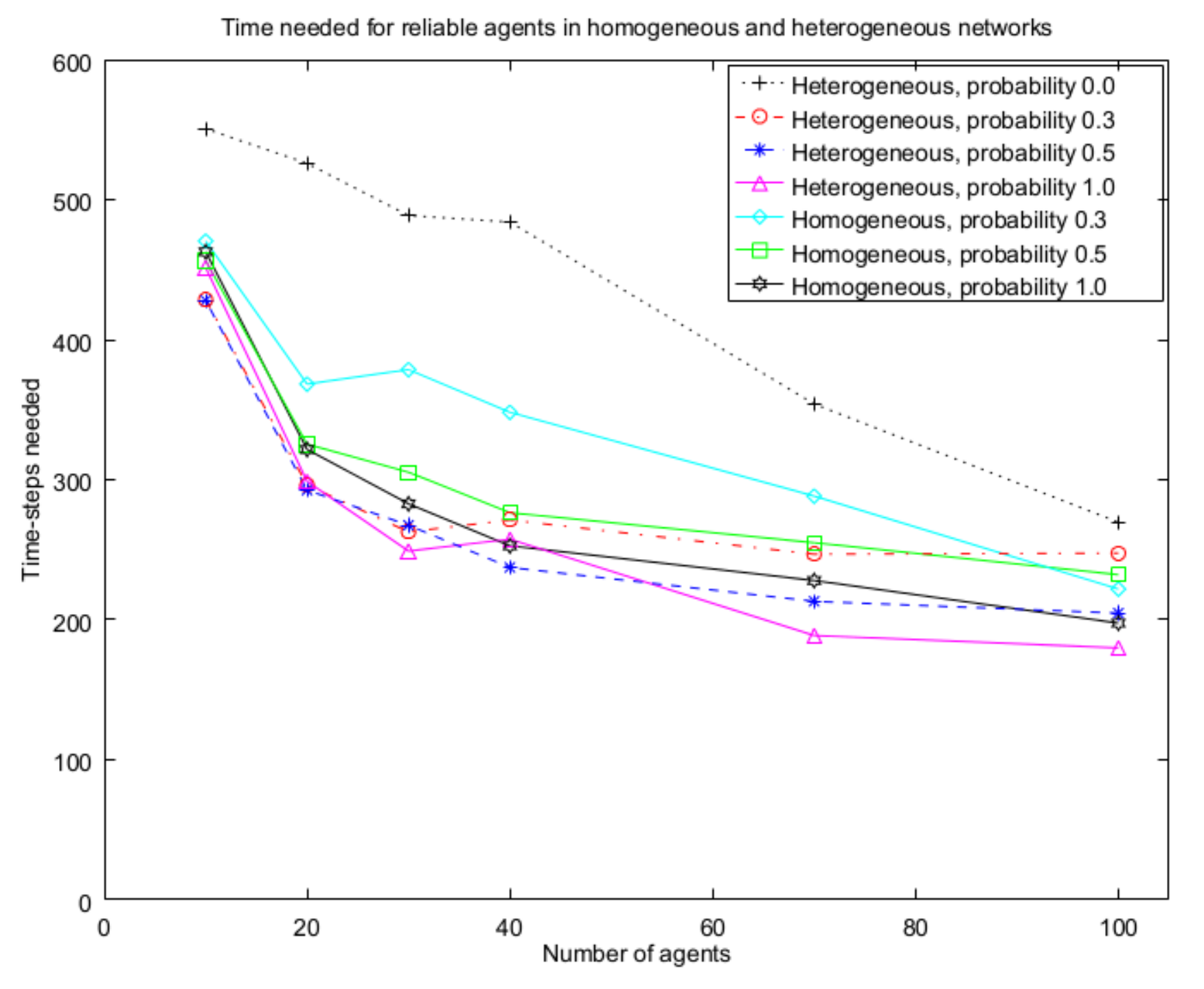}
\caption{Time needed}
\label{fig:combitime}
\end{figure}

\dd{In what follows we present the most significant results of our
  simulations. For each of the comparisons below we will vary only
  the parameter given in the respective paragraph title while
  keeping all other parameters fixed.

  \textbf{Increased information sharing.} Increased information
  sharing appears to be beneficial for both reliable and biased
  groups. On the one hand, homogeneous groups that share information
  with probability 1 perform faster, while being similarly successful
  as those with lower strict positive probabilities (SPP) of
  information sharing (Fig.\ \ref{fig:sucrelhomog} and
  \ref{fig:combitime}).
  } As expected, homogeneous groups
  that do not share information among each other perform the worst in
  terms of both, success and time. On the other hand, heterogeneous
  groups that share information are all similar for SPP in terms of
  time and success, (see Fig.~\ref{fig:sucrelheterog} and
  \ref{fig:combitime}), except for larger groups (70 and 100 agents)
  which are faster in case the probability of sharing information is
  1. Groups with probability 0 perform worse in both
  respects. 

  \textbf{Reliable vs.\ biased agents.} With respect to homogeneous
  groups, reliable agents appear to be more successful than the biased
  ones in case of SPP of
 information sharing, while sometimes
  being only slightly slower. With respect to heterogeneous groups,
  reliable and biased groups perform similarly in terms of both
  success and time.

  \textbf{Homogemous vs.\ heterogenous networks.} In case of 0.3 and
  0.5 probabilities of information sharing, heterogeneous networks
  tend to be faster than the homogeneous ones, while being similar in
  terms of success. In case the probability is 1, heterogeneous and
  homogeneous groups are similar in both respects (Fig.\
  \ref{fig:combitime}). In case of no information sharing
  heterogeneous groups are much more successful. These results hold
  for both reliable and biased agents.

\section{Discussion}
\label{sec:discussion}

In this section we will first compare our results with those obtained
by other ABMs of science, and then we will turn to a critical analysis
of some idealizations present in our model.

\new{Our finding that increased communication tends to be
  epistemically beneficial (or at least, not epistemically harmful)
  undermines the robustness of conclusions drawn from ABMs in
  \cite{zollman2007communication,zollman2010epistemic,grim2009threshold,grim2013scientific},
  under different modeling choices. As we argue below, there is no
  reason to assume that any of these ABMs represent scientific
  interaction more adequately than the presented ABM does. In
  addition, our finding that biased agents perform worse than reliable
  ones under conditions of an increased information flow shows that
  once we apply the notion of bias to the way in which agents share
  information (rather than to their confidence in the given theories
  with which they begin an inquiry, as done in
  \cite{zollman2010epistemic}), we get results that are contrary to
  \cite{zollman2010epistemic}.} 

\new{While} a number of open issues regarding the comparison of our
results with those obtained by means of other ABMs remain for future
research, we can already highlight the features of our model that
allow for a more adequate representation of scientific inquiry than
this has so far been done with ABMs. First of all, our model addresses
problematic features identified in other ABMs (see Section
\ref{sec:intro}), namely (i) the inadequate representation of heuristic
appraisal, as well as (ii) the inadequate representation of information
flow among scientists. Regarding (i),
scientists in our model are equipped with certain prospective
considerations, explained in Section \ref{sec:basic-behav-agents}.
In the current version of the model, such prospective considerations
play a role in methods that guide agents in their inquiry. It remains
a task for a future version of the model to also incorporate such
features into the evaluations performed by agents in view of which
they can judge how promising their theory is. Such an assessment would
more aptly capture heuristic appraisal, which informs scientists how
worthy of pursuit different theories are, rather than how confirmed
(or defensible) they are in view of the available
evidence. 

Regarding (ii), the notion of information flow is in contrast to many other
ABMs of science not just a simple update of information. Instead,
exchange of information is represented as argumentative, which means
that received information is critically assessed. Hence, when an
agent receives new information regarding a discovered argument, she
will assess it as acceptable (in case it can be defended in view of
her knowledge base) or as unacceptable (in case it cannot be
defended in view of her knowledge base).

Moreover, different information triggers different heuristic
activities on the side of the receiver, so that an agent who has
discovered an attack on her current argument tries to find its
defense. Finally, we acknowledge the fact that receiving information
(such as reading articles by other scientists) may be time costly.

Nevertheless, our ABM is still based on a number of idealizations, the
impact of which should be examined in future research. First,
heuristics of agents are highly simplified, including their search for
defense in view of discovered attacks. Even though agents may
recognize a defense in case it is located in their surroundings
\dd{(i.e.\ in one of the child-arguments of the given attacked
  argument, which an agent currently explores)}, a defense may be present at an entirely
different branch of the tree. Equipping agents with more insight into
where a defense may be, and thus representing a scientist as having
methods for finding solutions for the current anomalies of the theory,
is another task for future research.

Second, anomalies of a given theory are currently represented solely
as attacks from one of the rivaling theories. An improved version of
the model should allow for anomalies to be represented as
counter-evidence discovered only as a result of exploring the given
theory. Future versions of our model will include a more direct representation of evidence.

Third, what parameter settings for building landscapes in our model
are representative of specific types of scientific controversies is an
open question. 


In view of these remarks, it is important to interpret the results of
our runs cautiously when it comes to their relevance for the actual
scientific practice. Just like all other existing ABMs of science, our
model is still just a ``bookshelf model'', which means that it is
still too simplified to be fully informative of actual scientific
practices.\footnote{The importance of distinguishing between
  ``bookshelf models'' and those that are relevant for real world
  phenomena has been emphasized in \cite{pfleiderer2014chameleons} in
  the context of economic models. The same considerations apply to
  models of scientific inquiry.} Nevertheless, as we \dd{have argued},
the highly modular nature of our model together with its specific
design features makes it significantly closer to the aim of
representing scientific inquiry than the currently existing ABMs. As
such, our model offers a fruitful basis for further improvements,
which can provide insights into real world phenomena.


\new{Finally, let us compare our model with Gabbriellini and Torroni's
  (G\&T) ABM \cite{gabbriellini2014new}. Their aim is to study
  polarization effects among communicating agents, for instance, in
  online debates. Similarly to our approach, their model is based on
  an abstract argumentation framework. Agents start with an individual
  partial knowledge of the given framework and enhance their knowledge
  by means of communication. Since G\&T do not model inquiry, their
  agents cannot discover new parts of the graph by means of
  `investigating' arguments. Rather, they exchange information by
  engaging in a dialogue modeled after Mercier \& Sperber's
  argumentative theory of reasoning~\cite{gabbriellini2013ms}. This
  way, agents may learn about new arguments and attacks but also
  remove attack relations. Whether new information is incorporated in
  the knowledge of an agent depends on the trust relation between the
  discussants. The beliefs of agents are represented by applying
  Dung-style admissibility-based semantics to the known part of the
  argumentation framework of an agent. This is quite different from
  our model where the underlying graph topology is given by several
  discovery trees of arguments representing scientific theories and
  attacks between them. This additional structure of the argumentation
  graph is essential since we do not model the agents' beliefs in
  individual arguments but rather evaluative stances of agents that
  inform their practical decision of which theory to work on. While an
  admissibility-based semantics would lead to extensions that feature
  unproblematic sets of arguments from different theories (ones that
  form conflict-free and fully defensible sets), in our approach
  agents pick theories to work on. For this, they compare the merits
  of the given theories, pick the one that is `most' defended (where
  typically no theory is fully defended before the end of a run), and
  employ heuristic behavior to tackle open problems of theories (represented by
  incoming attacks). It will be the topic of future
  research to include dialogue protocols that are relevant for
  scientific communication, such as information-seeking, inquiry and
  deliberation dialogues~\cite{walton1995commitment}. }

\section{Outlook and conclusion}
\label{sec:outlook-conclusion}

In this paper we have presented an ABM of scientific inquiry which
makes use of abstract argumentation, aiming to model the argumentative
nature of scientific inquiry. The presented version of our model is
designed to tackle the question how different degrees of 
information flow among scientists affects the efficiency of their
knowledge acquisition. Our results suggest that an increased
information sharing is epistemically beneficial, which undermines the
robustness of contrary results obtained by previous ABMs of
science, under different modeling choices. While we have argued that
our model represents scientific inquiry more adequately than the
previous ABMs of science, we have also emphasized a number of issues
that remain to be tackled in future research. 
We will conclude the paper by showing the fruitfulness of our ABM for
the investigation of related questions concerning social aspects of
scientific inquiry.

First, our model can be enhanced with different types of research
behavior, such as ``mavericks'' and ``followers'', introduced in
\cite{weisberg2009epistemic}. 
Next, the
model can be enhanced with other aspects of scientific inquiry, such
as an explanatory relation and a set of explananda
\cite{Seselja_Strasser_11}. This would allow for an investigation of
different evaluative procedures which agents perform when selecting
their preferred theory (e.g.\ in addition to the degree of
defensibility, agents can take into account how much their current
theory explains, or how well it is supported by evidence).
Furthermore, a number of enhancements available from the literature on
AFs, such as probabilistic semantics \cite{thimm2012probabilistic},
values \cite{journals/corr/cs-AI-0207059}, etc.\ can be introduced in
future versions of our ABM. 

\paragraph{Acknowledgements}
The research of Annemarie Borg and Christian Stra\ss er was supported by a Sofja Kovalevskaja award of the Alexander von Humboldt-Foundation, funded by the German Ministry for Education and Research.

\bibliographystyle{splncs03}
\bibliography{acd}

\end{document}